\documentclass[twoside]{ilcws08}
\usepackage[latin1]{inputenc}
\usepackage{graphicx}
\usepackage{wrapfig,rotating}
\usepackage{amssymb,amsmath,array}

\begin{document}
\title{
Development of Vertically Integrated Circuits for ILC Vertex Detectors} 
\author{Ronald Lipton \\
for the Fermilab Pixel R\&D Group
\vspace{.3cm}\\
Fermilab \\
P.O. Box 500, Batavia, Illinois, USA
}

\maketitle

\begin{abstract}
We report on studies of vertically interconnected electronics (3D) performed by the 
Fermilab pixel group over the past two years.  These studies include exploration of 
 interconnect technology, backside thinning and laser 
annealing, the production of the first 3D chip for particle physics, the VIP, and 
plans for a commercial two-tier 3D fabrication run.  Studies of Direct bond Interconnect 
(DBI) oxide bonding and Silicon-on-Insulator based technologies are presented in other talks in this 
conference.
\end{abstract}

\section{3D electronics}
3D electronics is generally defined as consisting of multiple layers of electronics, 
thinned, bonded and interconnected to form a monolithic circuit.  This technology 
has become an area of intense focus in the electronics industry as a way to 
improve circuit performance without the expense and complexity of smaller feature size~\cite{IBM}~\cite{3Dref}. 
3D electronics provides the ability to integrate heterogeneous technologies, reduce 
interconnect lengths, and expand bus width.  These technologies are particularly interesting 
for pixel detectors, where they offer new techniques for integrating sensors and electronics, 
and provide substantially more processing power per pixel than conventional technologies.

Fabrication of a 3D stack depends on the development of several techniques including:
\begin{itemize}
\item Bonding between layers including oxide to oxide fusion, copper/tin bonding,
copper/copper bonding, polymer/adhesive bonding
\item Wafer thinning using grinding, lapping, etching, and Chemical Mechanical Polishing (CMP)
\item Through wafer via formation and metalization either with isolation using Through Silicon Vias (TSVs)
or without isolation as in Silicon-on-Insulator (SOI) devices.
\item High precision alignment
\end{itemize}
The technology can also be separated into techniques which form the via before 
3D integration is performed (via first) and those which form vias afterwards (via last). 

\section{VIP Chip}
The VIP chip~\cite{VIP} was intended as a demonstration of 3D technology as applied to an ILC vertex detector.
It was fabricated in  0.18 micron SOI CMOS technology by MIT-Lincoln Labs as part of a DARPA-sponsored 
multiproject run.  The chip consists of three tiers of electronics interconnected by 3D vias. The MIT-LL technology 
utilizes oxide bonding to bond tiers together. This wafer bonding technology mates activated, planarized silicon
oxide surfaces to form a robust inter-wafer bond~\cite{MIT-LL}.  The top "handle" silicon in the bonded wafer stack is thinned to 
$70 \mu m$ and the remainder of the silicon is etched away to expose the oxide surface.  Vias are then formed 
by etching thorough the insulating oxide to contact internal metalization layers. The process is repeated to form the 
required number of tiers.

\begin{figure}[htb]
\includegraphics[width=50mm]{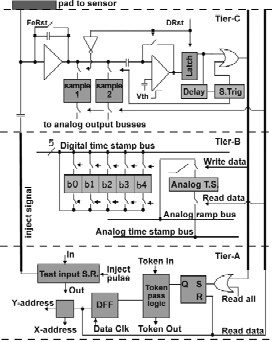}
\includegraphics[width=90mm]{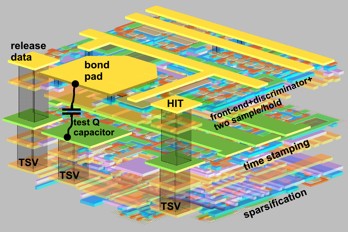}
\caption{Schematic of the VIP chip showing the three tiers of electronics. The drawing on the right shows the 
metalization for the three layers. }
\label{VIP1}
\end{figure}

The VIP incorporates an amplifier/disriminator with double correlated sample and hold in tier three, 
both a 5-bit digital and an analog time stamp in tier 2, and sparsification and digital logic in tier one,
all within a $20 \mu m$ square pixel.  In ILC operation time stamped hits are stored within the pixel 
during the bunch train.  The chip utilizes part of  the 199 ms period between trains for readout, while the front end current 
is reduced to save power.  The front end is designed to consume less than 4 mW/mm$^2 \times f$ where $f$ is the 
front end duty factor.  A token passing readout scheme stores addresses on the periphery, 
minimizing the logic on the pixel. Figure~\ref{VIP1} shows a schematic of the chip as well as a visualization 
of the metal layers.  

\begin{figure}[htb]
\includegraphics[width=130mm]{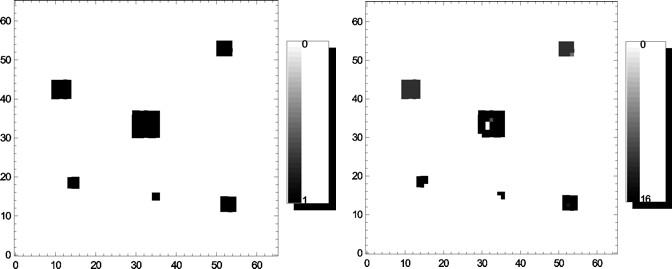}
\caption{Pattern of 119 pixels injected with a test charge (left) and read out in the subsequent sparse scan.}
\label{VIP_qinj}
\end{figure}

The chip was submitted in October 2006 and received in October 2007.  Initial tests showed that the 
overall yield was low, with only a few chips showing the ability to propagate the readout token through the 
full 64 x 64 matrix. The single best-performing chip was selected for full testing. Figure~\ref{VIP_qinj}   shows the results 
of the most complete system test, where a pattern of test pulses are injected into the front-end amplifiers,
and a sparse scan is performed, reading out those channels where the discriminator fired and latched the 
readout flag.  

Our testing has demonstrated the basic functionality of the chip including propagation of the readout token,
threshold scans, input test charge scans, verification of digital and analog time stamping, full sparsified data readout,
and fixed pattern and temporal noise measurements.  
No problems could be found associated with the 3D vias between tiers.  Although the chip was fully functional, 
we were not fully satisfied with yield and performance.  Performance problems stemmed from large leakage currents 
in the protection diodes and transistors, and poor matching of current mirrors.  Many of these issues 
can be traced to the sensitivity of mixed mode designs in fully depleted SOI to the transistor environment and process 
variations~\cite{IBMQ}~\cite{SOIQ}. Intrinsic SOI process 
problems were exacerbated by our aggressive design, which made extensive use of minimum feature size transistors 
and dynamic logic, which is sensitive to transistor leakage current.

An new version of the chip, the VIP2, was submitted to the third DARPA sponsored 3D multiproject run in October 2008.
As a result of useful interaction with MIT-LL on SOI  analog design the overall quality should be considerably improved.
Changes to the chip include:
\begin{itemize} 
\item Different power and grounding layout
\item Larger transistor sizes ($0.18 \to 0.45 \mu m$) equivalent feature size
\item Larger pixels (30 x 30 microns)
\item Redundant vias and larger traces in critical paths
\item Redesign of current mirrors to reduce thermal effects
\item Removal of dynamic logic due to leakage current problems
\end{itemize}
Changes were also made to improve the overall functionality including increasing the digital time stamp from 
5 to 7 bits.
We expect that the changes will lead to a much more reliable chip which can be bonded to sensors for test beam studies.

\section{Commercial 3D Technologies}
An R\&D process, such as the one provided by  MIT-LL has disadvantages of long turn-around time and 
process uncertainties.   We are now exploring alternative 3D processes implemented as part of a high volume 
commercial process. Tezzaron (Naperville Ill) has developed a 3D technology implemented in the high volume 
0.13 micron Chartered (Singapore) process~\cite{3Dref}.  This is a "via first" process where through-silicon "supercontacts" are formed 
after transistor fabrication but before any metalization processing.  The 6 micron deep by 1 micron diameter supercontacts are 
filled with tungsten at the same time as the transistor contacts are formed.  Wafers are finished normally, however 
there is a top layer of thin patterned copper which forms both the bond between wafers and the wafer-to-wafer electrical 
interconnection. Wafers are then bonded face-to-face with moderate pressure and temperature.  Silicon on the top 
wafer is ground down to the supercontacts and the contacts are metalized to form either external bond pads 
or to provide connections to the next tier. 

Fermilab is organizing a multiproject run in the Tezzaron/Chartered process with submission expected in Spring of 
2009.  The run includes designs from 13 institutes.  The Fermilab designs will include a two tier version of the VIP
chip, the VIP2b, as well as test devices for the CMS upgrade and X-ray imaging. Standard commercial CMOS 
should provide a reliable process with low noise, multiple transistor options, better rad hardness, 
less wasted via area, faster turn-around, as well as the availability of full wafers for sensor integration.

\section{Sensor Integration}
The bonding, thinning and lithography process used to build 3D tiers  can also be used for sensor 
integration with readout.   Both Tezzaron and Ziptronix provide 3D processes, based on copper-copper and oxide 
bonding respectively, which can be used to  include sensors as a base tier in a 3D stack.  3D processes 
offer finer pitch and more robust mechanical interconnection than is available in solder-based bump bonding.
The high planarity and strong interlayer bonds allow bonded readout ICs to be thinned to 25 $\mu m$ or less, which 
can provide access to integrated through-silicon vias. 

We have explored two sensor bonding techniques: Cu-Sn and DBI oxide bonding.  The DBI technique is described in another 
contribution to this conference~\cite{DBI_ye}.  We have contracted with RTI to explore Cu-Sn bonding with sparse contacts to 
minimize contact mass. These technologies do not have the contact bridging problems that limit pitch for 
solder bumps.  However they are not self-aligning which requires special care in aligning the bonded surfaces.
In that study successful Cu and Cu-Sn bump structures were fabricated that were 
compatible with $20 \mu m$ I/O pitch. Electrical tests indicated that the bonding yield was  $> 99$ \%
for both metallurgies.  Samples were also destructively tested to evaluate bond strength.  In both cases the
bond strength is considerably higher than comparable solder bump arrays~\cite{Huffman}.



\section{Thinning and Laser Annealing}
Wafer thinning is an important part of 3D technology, and the ability to process and handle 
thinned silicon is crucial to the goal of constructing a very low mass vertex detector. In 
some 3D and SOI technologies it will be important to thin the devices after topside processing.  After 
thinning a backside contact ohmic contact must also be formed.  The contact is usually fabricated by a 
high temperature anneal of an ion implantation.  This high temperature step is unacceptable for 
fully processed electronics, where the temperature must be kept below ~450 degrees C. to protect 
the topside metalization.  

We have developed a thinning/implantation/annealing process which limits the maximum temperature 
of the topside to below 100 deg C.  The wafer is first bonded to a pyrex carrier using a 3M UV release 
adhesive designed for wafer thinning applications.  The wafer is then thinned and polished using 
standard techniques.  The bonded wafer is ion implanted, taking care to ground the silicon edges and 
controlling the implantation rate to limit the temperature.   The implantation is then annealed using a 
eximer laser system which melts the silicon locally on the backside to a depth of $300$ nm, keeping 
the topside close to room temperature.

\begin{figure}[htb]
\includegraphics[trim = 0mm 20mm 0mm 0mm, clip, width=80mm]{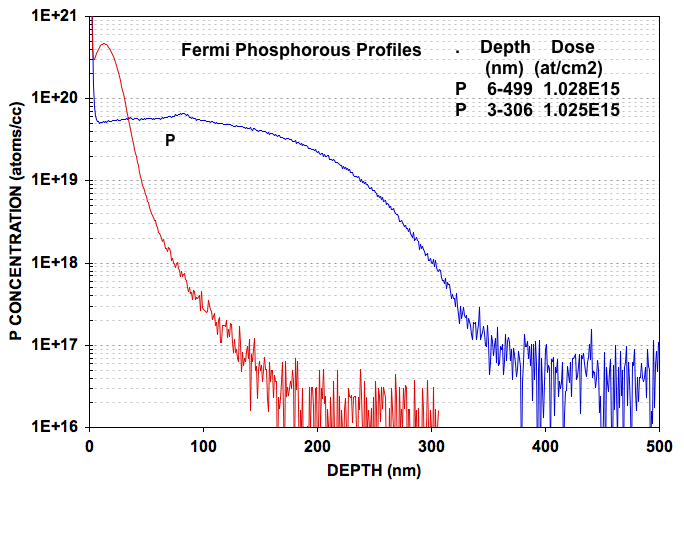}

\caption{Secondary Ion Mass 
Spectrometry (SIMS) phosphorus dose profile of  strip  detector before (red) and 
after (blue) laser annealing.}
\label{sims}
\end{figure}

Initial studies were performed using individual strip sensors, which were thinned to 
remove the backside ohmic implant, re-implanted and laser annealed.  All of these 
devices showed acceptable performance at depletion, with some variation of leakage 
current depending on laser dose and annealing environment.  A scan of the 
dopant concentration before and after the annealing process is shown in Figure~\ref{sims}.
This work was followed by studies using 6" test wafers donated by 
Micron Semiconductor.  These wafers were thinned to 50 microns on the pyrex handle, 
implanted and laser annealed at MIT-Lincoln Labs.  Depletion voltage was reduced from 80 to 
2.5 volts with acceptable leakage current. Work is ongoing at Cornell to determine the optimal 
implantation and annealing parameters.

\section{Ladder design}
\begin{figure}[htb]
\includegraphics[width=70mm]{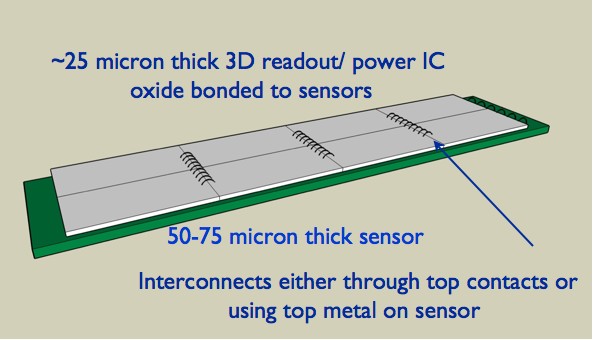}
\includegraphics[trim = 0mm 30mm 0mm 80mm, clip, width=80mm]{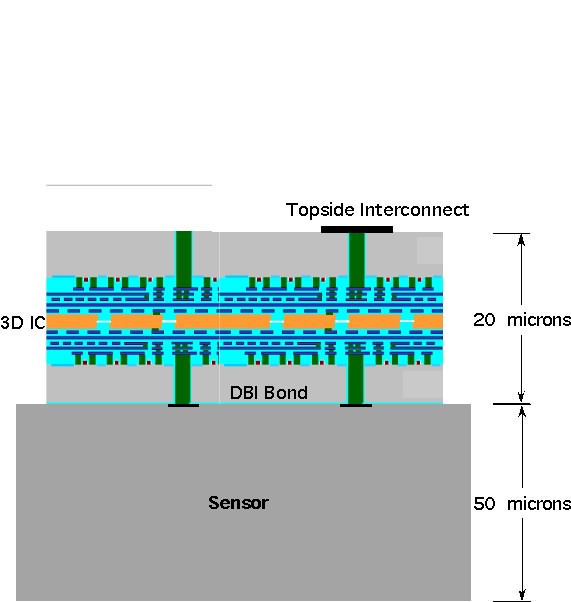}
\caption{Conceptual drawing of a thinned ladder(left) based on 3D interconnects (right). Readout chips could be bonded to 
the sensor wafer using an oxide bonding process (DBI) then thinned to $\approx 25 \mu m$ to expose through-silicon vias
to provide interconnections.
The sensor wafer could be pre-thinned and mounted on a handle wafer during the DBI processing.}
\label{3D_ladder}
\end{figure}

3D technologies provide  the ability to construct a low mass, dense, tiled array of chips, which can be used to fabricate 
ladder and disk planes for the ILC.  
Figure~\ref{3D_ladder} shows an example of such a structure.
Multi-tier readout ICs are fabricated utilizing through-silicon vias. These ICs are bonded to an independently 
fabricated sensor wafer with a fine pitch technology such as DBI or cu-cu.   Once 
bonded, the Readout ICs are thinned to reveal the topside TSVs. Topside interconnections are patterned 
using standard lithography
and wirebond or other contacts are made.  Connections to external power and signal cables could 
be made at the ends of the sensor, which could have the appropriate interconnection patterns.
This technique has a number of advantages:
\begin{itemize}
\item The sensor can be a fully depleted detector with charge collection by drift rather than diffusion.
\item The sensor wafer serves as a base for the ROICs, obviating the need for reticle stitching.
\item The 3D ROICs can include power control tiers.
\item Known good ROIC die can be used.
\end {itemize}
Final sensor thinning would have to occur after the topside processing.  This could be done by 
backgrinding a sensor with an imbedded ohmic contact, either as part of an epitaxial stack or 
using the SOI technique demonstrated by the Max Plank Institute~\cite{Fischer:2007zzm}. This would avoid the additional 
implant and laser annealing steps needed if the backside were not pre-processed.

\section{Conclusions}
Fermilab has produced and tested the VIP, the first 3D chip designed for 
particle physics applications.  The chip demonstrated the required functionality 
but suffered from low yield and compromised performance.
An improved version of the chip has been submitted to MIT-LL.  The VIP2b, a two-tier 0.13 micron  
CMOS chip implemented in the Tezzaron 3D process, will be submitted this spring.  
This submission will extend the development of this technology to 
applications at super-LHC and in x-ray imaging.
We are continuing to 
develop wafer thinning, interconnection, and post-processing technologies 
aimed at demonstrating the ability to build precise, low mass, low power vertex 
detector systems.


\begin{footnotesize}



%

\end{footnotesize}


\end{document}